# Development and evaluation of a tutorial to improve Students' Understanding of a lock-in amplifier

Seth DeVore, Alexandre Gauthier, Jeremy Levy, and Chandralekha Singh

*Department of Physics and Astronomy, University of Pittsburgh, Pittsburgh, PA 15260*

**Abstract:** A lock-in amplifier is a versatile instrument frequently used in physics research. However, many students struggle with the basic operating principles of a lock-in amplifier which can lead to a variety of difficulties. To improve students' understanding, we have been developing and evaluating a research-based tutorial which makes use of a computer simulation of a lock-in amplifier. The tutorial is based on a field-tested approach in which students realize their difficulties after predicting the outcome of simulated experiments involving a lock-in amplifier and check their predictions using the simulated lock-in amplifier. Then, the tutorial provides guidance and strives to help students develop a coherent understanding of the basics of a lock-in amplifier. The tutorial development involved interviews with physics faculty members and graduate students and iteration of many versions of the tutorial with professors and graduate students. The student difficulties with lock-in amplifiers and the development and assessment of the research-based tutorial to help students develop a functional understanding of this device are discussed.

**Keywords:** lock-in amplifier, tutorial, think-aloud protocol, physics education research

**PACS: 01.40Fk, 01.40.gb, 01.40G-**

## INTRODUCTION

The physics graduate student population is one of the most under-addressed groups in Physics Education Research (PER). While the group represents students with many years of directed study within the field of physics, it is difficult to find research-based tools with the express intent of addressing student difficulties of physics graduate students and providing scaffolding support for the continued education of those at the highest tiers of physics learning. However, not only can PER help physics graduate students develop functional understanding of physics in the graduate level courses, PER-based tools can also be helpful for them once they move beyond the classroom and into their roles as researchers. While these students have been through the entire undergraduate physics curriculum, they can still benefit from research-based intervention to help them learn, e.g., with tutorials based upon physics education research.

To aid in physics graduate students' entry into the lab setting which requires the use of a lock-in amplifier, we developed and evaluated a research-based tutorial to help students develop a functional understanding of this device. The tutorial strives to help students integrate conceptual and quantitative aspects of how this device works so that they can use the device effectively and trouble-shoot problems that may arise.

The lock-in amplifier (lock-in) is an instrument used extensively in lab research, especially in condensed matter physics [1-6]. The most common use of a lock-in in laboratory research is to measure small signals that are synchronous with an external reference, often in the presence of large background signals (or "noise"). However, many students who use this instrument for their research have only a limited understanding of the operation of lock-ins. Often, the lock-in is used as a "black box" to find the amplitude of a signal at a given frequency without understanding the instrument's internal workings. According to the experimental condensed matter physics faculty members interviewed by us during this research-project, improper or inefficient use of the lock-in and misinterpretation of data obtained from the device are quite common among the students in their research lab. Additionally, according to these faculty advisors, the lack of understanding can often result in students being unable to troubleshoot and modify the experimental setup when faced with anomalous results.

Computer-based and web-based learning tools are becoming increasingly common to aid in learning across many science and engineering fields [7-18]. These tools must be developed using a research-based approach to ensure that they are effective and suit both the level and the prior experience of the students they are intended for [19,20]. Here, we discuss the development and evaluation of the lock-in tutorial which makes use of a simulation of the lock-in device to ease the transition of those who are just beginning to use this instrument in their research in the lab setting, as well as to provide a firmer foundation for those who already use lock-ins in their research. Our goal in developing this research-based tutorial was to help students understand the basics of the lock-in functions. In particular, the tutorial strives to help students who use (or plan to use) lock-ins in their research understand more deeply how the input parameters in diverse situations affect the output and why so that they learn to predict the output of the device for a set of input parameters. In particular, by merging conceptual and mathematical aspects of the operation of the instrument, the tutorial strives to help students learn the relationship between the input parameters and expected outputs so that they are able to troubleshoot unexpected outputs in their lab work if the output does not seem consistent with the input they used (e.g., due to various types of noise or inappropriate setting of the time-constant for a given experiment).

In the following sections, we will begin with a description of the idealized lock-in in which we will summarize the mathematical preliminaries required because integrating conceptual and quantitative aspects of the device is critical for a robust understanding of the device. This section is followed by the methodology used in the development of the tutorial including methods for investigating student difficulties in using this device. We then summarize the structure of the tutorial including the pretest and posttest and the associated simulation. This is followed by an examination of the student difficulties that came to light in our initial investigation and were emergent from think-aloud interviews in which students made use of early versions of the tutorial. Investigation of these student difficulties was critical for addressing them using a guided approach to learning used in the lock-in tutorial. Finally, we discuss the evaluation of the tutorial using a pretest and posttest.

## THE IDEAL LOCK-IN AMPLIFIER

Throughout this paper, as in the tutorial, we will treat the lock-in as an idealized version of the instrument. Here we assume that the signal of interest is centered on a frequency $f_S$ which is present in the input signal. In general, it will not be a pure frequency since the amplitude can change, and amplitude modulation leads to sidebands that surround the central frequency. In the case in which there is no amplitude modulation introduced into the signal we will treat this frequency as a pure frequency. Throughout this tutorial we make use of both idealized pure frequencies and input signals which are undergoing amplitude modulation to ensure that students will gain experience with both treatments of the lock-in. To separate the signal of interest from unwanted noise, a reference signal is defined. The reference signal is selected to have a unit amplitude (for convenience). The (idealized) single-frequency input signal is first pre-amplified by a factor $g$, to give

$$V_I = gA_S \cos(2\pi f_S t + \varphi). \tag{1}$$

This amplified signal is then multiplied (or "mixed") by a reference for the *x* and *y* channels of the lock-in:

$$v_{RX} = \cos(2\pi f_R t) \quad (2a) \qquad\qquad v_{RY} = \sin(2\pi f_R t) \quad (2b)$$

to form the "unfiltered" *x* and *y*-channel outputs of the lock-in:

$$V_{MX}(t) = V_I(t)v_{RX}(t) \quad (3a) \qquad\qquad V_{MY}(t) = V_I(t)v_{RY}(t) \quad (3b)$$

Here, $\varphi$ is the phase of the input signal of frequency $f_S$ with respect to the reference signal, and $A_S$ is the amplitude of the input signal with frequency $f_S$. To understand the effect of the mixer, we rely on two trigonometric identities:

$$\cos(a)\cos(b) = \tfrac{1}{2}[\cos(a+b) + \cos(a-b)] \tag{4a}$$
$$\cos(a)\sin(b) = \tfrac{1}{2}[\sin(a+b) - \sin(a-b)]. \tag{4b}$$

Application of these identities yields:

$$V_{MX} = V_I v_{RX} = \tfrac{1}{2}gA_S[\cos(2\pi(f_S - f_R)t + \varphi) + \cos(2\pi(f_S + f_R)t + \varphi)] \tag{5a}$$
$$V_{MY} = V_I v_{RY} = \tfrac{1}{2}gA_S[\sin(2\pi(f_S - f_R)t + \varphi) - \sin(2\pi(f_S + f_R)t + \varphi)] \tag{5b}$$

In most experimental situations, the signal is close in frequency to the reference: $f_S - f_R \ll f_R$ and $f_S + f_R \approx 2f_R$. For the case $f_S = f_R$, we have $f_S - f_R = 0$ and $f_S + f_R = 2f_R$, and the unfiltered *x*-channel and *y*-channel outputs contain a rapidly oscillating term superimposed on a time-independent one:

$$V_{MX} = V_I v_{RX} = \tfrac{1}{2} g A_S [\cos(\varphi) + \cos(2\pi(2f_R)t + \varphi)] \tag{6a}$$
$$V_{MY} = V_I v_{RY} = \tfrac{1}{2} g A_S [\sin(\varphi) - \sin(2\pi(2f_R)t + \varphi)] \tag{6b}$$

Finally, $V_{MX}$ and $V_{MY}$ are each fed through a low-pass filter with a "time constant" $\tau = 1/(2\pi f_c)$ where $f_c$ is the "cutoff" or "corner" frequency of the filter; the filter "rolloff" is most commonly chosen to be one of four values (6 dB/octave, 12 dB/octave, 18 dB/octave, and 24 dB/octave). The values selected for both the time constant and the rolloff should be chosen carefully based upon the nature of the experiment. As a "rule of thumb", the $6n$ dB/octave filter preserves signals with frequency $f \ll f_c$, while attenuating signals with $f \gg f_c$ according to a power law $f^{-n}$ for (e.g., $\propto f^{-2}$ for 12 dB/octave filters). The resulting filtered outputs are defined as $V_{OutX}$ and $V_{OutY}$ in the tutorial. In the idealized version of the case in which $f_R = f_S$, the time constant should be selected such that the low-pass filter attenuates the second-harmonic ($2f_R$) term from both $V_{MX}$ and $V_{MY}$ resulting in a time-independent output signal. The relationship between $V_{OutX}$ and $V_{OutY}$ and the magnitude and phase of the input signal as initially defined is given by the familiar trigonometric identities:

$$V_{OutX} = \tfrac{1}{2} g A_S \cos \varphi \tag{7a} \qquad V_{OutY} = \tfrac{1}{2} g A_S \sin \varphi \tag{7b}$$

or

$$A_S = (2/g)\sqrt{V_{OutX}^2 + V_{OutY}^2} \tag{8a} \qquad \varphi = \tan^{-1}(V_{OutY}/V_{OutX}) \tag{8b}$$

As noted earlier, the most common use of a lock-in in laboratory research is to measure small signals that are synchronous with an external reference, often in the presence of large background signals or noise. For this class of measurements, the reference frequency is set equal to the signal frequency ($f_R = f_S$). However, there are applications in which the two frequencies would not be the same; alternately, when changing a parameter that affects the signal, one is effectively modulating the signal in time. This type of "amplitude modulation" will produce sidebands around $f_S$ which must be passed with acceptably low attenuation, while making sure that the signal is not 'flooded' with background noise. While the lock-in is most commonly used in the ideal case ($f_R = f_S$ with the $2f_R$ strongly attenuated) with no other superimposed signals, there are many instances in which unwanted signals can interfere with measurement. Power line noise (i.e. the 50 Hz or 60 Hz frequency introduced by ac electrical power) can often superimpose large sinusoidal signals over the desired measurements. Helping students develop a functional understanding of the common uses (and misuses) of lock-in amplifiers represents the primary goal of the lock-in tutorial.

## METHODOLOGY

Before the development of the tutorial and the associated pretest and posttest (pretest is administered before the tutorial and posttest after it), informal discussions with faculty members who conduct research in experimental condensed matter physics using the lock-in were carried out. These discussions provided information about the fact that even the graduate student who have been using this device for some time have many common difficulties in using this device effectively. Therefore, we decided to start the development of the tutorial by conducting audio-recorded interviews with several faculty members. We individually interviewed five physics professors at the University of Pittsburgh who conduct research in condensed matter physics and commonly work with graduate students who use lock-ins in their research. A typical interview time was 60-90 minutes, during which each professor was also asked to articulate what they expected their students to know about lock-ins, what was the goal of the professor's experimental research projects utilizing lock-ins, how lock-ins are useful in the broader framework of their research and what were the common difficulties their students had with the device. While two co-authors of this manuscript conduct research in experimental condensed-matter physics, the other two had an opportunity to play with an actual lock-in in a research lab. Using the feedback from professors as a guide along with a cognitive task analysis of the underlying knowledge involved in the operation of a lock-in, we developed a preliminary tutorial along with a pretest and posttest (to be given before and after the tutorial, respectively). The development of the initial version of the tutorial was accompanied by the development of a lock-in simulation. The simulation is integral to the tutorial and was built into the structure of the tutorial. The simulation was integrated with the tutorial with the intent to give students an opportunity to experience the operation of the device from anywhere to develop a functional understanding about it using a proxy for the actual device.

We then interviewed graduate students who made use of the most up to date version of the tutorial using a think-aloud protocol to better understand their difficulties and to fine-tune the tutorial as well as its associated pretest, posttest and simulation. In these semi-structured interviews, students were asked to talk aloud while they worked through the pretest, tutorial and the posttest. The interviewer tried not to disturb students' thought processes while they answered the questions except to encourage them to keep talking if they became quiet for a long time. Later, the interviewer asked students for clarification of points they had not made clear earlier in order to understand their thought processes better. These interviews were semi-structured in that some of these questions were planned out ahead of time while others were emergent queries based upon a particular student's responses to questions during an interview.

The tutorial (along with the pretest and posttest) was iteratively refined over 30 times, based upon feedback from graduate students and professors. After the tutorial and its supplementary material underwent this process of revision and fine-tuning, it was administered to 21 additional physics graduate students who had not been involved in the development phase of the tutorial. These students ranged in experience from those who had been introduced to the basics of the lock-in in their research group but had not made use of one for their own research to those with extensive experience with the lock-in and either concurrently used a lock-in for their research or had made extensive use of one in the past.

To evaluate the effectiveness of the tutorial, a pretest and posttest were developed. A generalized grading rubric (summary available in Table 1) was developed that is capable of grading questions with all of the possible outcomes that could result when solving each of the problems on the pretest and posttest involving the operation of the lock-in. To this end, three researchers jointly deliberated a series of six classes of rubrics that could be used in the scoring of all problems present in all versions of the pretest and posttest. Three rubrics were required to cover all possible output signals present in the pretest and posttest that could be achieved with either zero or nonzero amplitude DC components and zero or nonzero amplitude time varying components, and three were required for possible input signals with either one frequency, one frequency with a nonzero phase angle, or two frequencies present in the input signal. The final version of the rubric resulted in better than 90% inter-rater reliability between two researchers for the scoring of performance on the pretest and posttest questions.

**TABLE 1.** Summary of the grading rubrics used for the pretest and posttest questions discussed. Rubrics 1 through 3 are used to grade problems in which the students are asked to sketch the output signal for a given set of input parameters (output problems) and Rubrics 4 through 6 are used to grade problems in which the students are asked to provide the input parameters for a given output (input problems). Case 1 problems are graded by rubric 1 for output problems and rubric 4 for input problems. Case 2 problems are graded by rubric 2 for output problems and rubric 4 for input problems. Case 3 problems are graded by rubric 1 for output problems and rubric 5 for input problems. Case 4 problems are graded by rubric 3 for output problems and rubric 4 for input problems. Case 5 problems are graded by either rubric 1 or rubric 3 for output problems, depending on the time constant and amplitude modulation frequency given in the problem. There are no Case 5 problems in which output is provided and students are asked for input parameters. Case 6 problems are graded by rubric 3 for output problems and rubric 6 for input problems.

| Rubric 1: Correct answer should have no frequency and non-zero DC offset present in the output signal. | | | | |
|---|---|---|---|---|
| Non-zero frequency present? | Yes | 0 points | | |
| | No | Magnitudes of DC components (x,y) are? | Correct | 10 points |
| | | | Incorrect but non-zero | 5 points |
| | | | Zero | 0 points |
| Rubric 2: Correct answer should have a non-zero frequency and zero DC offset present in the output signal. | | | | |
| Non-zero frequency present? | Yes | Magnitudes of both DC components (x,y) are zero? | Yes | 5 points | +2.5 points for correct frequency |
| | | | No | 2.5 points | +2.5 points for correct amplitude |
| | No | 0 points | | |

| Rubric 3: Correct answer should have a non-zero frequency and non-zero DC offset present in the output signal. | | | | | |
|---|---|---|---|---|---|
| Non-zero frequency present? | Yes | Magnitudes of both DC components (x,y) are correct? | Yes | 5 points | +2.5 points for correct frequency +2.5 points for correct amplitude |
| | | | No | 2.5 points | |
| | No | Magnitudes of both DC components (x,y) are correct? | Yes | 2.5 points | |
| | | | No | 0 points | |
| Rubric 4: Correct answer should have one frequency in the input signal. | | | | | |
| Frequency Correct? | Yes | Amplitude Correct? | Yes | 10 points | Divide score by 2 if a second input signal is given that would result in a different output signal |
| | | | No | 5 points | |
| | No | Amplitude Correct? | Yes | 5 points | |
| | | | No | 0 points | |
| Rubric 5: Correct answer should have one frequency in the input signal with a phase angle. | | | | | |
| Frequency Correct? | Yes | Amplitude Correct? | Yes | 5 points | +5 points for correct phase angle Divide score by 2 if a second input signal is given that would result in a different output signal |
| | | | No | 2.5 points | |
| | No | Amplitude Correct? | Yes | 2.5 points | |
| | | | No | 0 points | |
| Rubric 6: Correct answer should have two frequencies in the input signal. | | | | | |
| Frequency Correct? | Yes | Amplitude Correct? | Yes | 5 points | Use the rubric for each of the two frequencies that should appear in the input signal and sum the points |
| | | | No | 2.5 points | |
| | No | Amplitude Correct? | Yes | 2.5 points | |
| | | | No | 0 points | |

# THE TUTORIAL STRUCTURE

The most up to date version of the tutorial, simulation and the associated pretest and posttest are available for download on ComPADRE [21]. As noted earlier, students first take a pretest before working on the tutorial. The pretest takes the form of a short quiz comprised of questions related to the function of a lock-in. These questions are posed making use of the interface of the simulation which was developed alongside the tutorial. For each of the questions, the student is provided with a screenshot of the simulation's interface with either the output erased or certain components of the input signal removed and the student is asked to provide the missing information (as seen in the pretest and posttest [21]). Students are provided with a sheet that contains supplementary information to aid them as they work through this pretest. The supplementary information provided contains a description of the simulation interface and explanation of how each of the controls modify the input signal, reference signal and low-pass filter. Additionally, students are provided with several graphs that describe the effect of the low pass filter, a diagram of the dual channel lock-in and a description of the effect of amplitude modulation in this simulation. This supplementary information was developed to ensure that student performance on the pretest was not the result of confusion about the form in which the questions were asked or a lack of understanding of some of the terms and information used in these pretest questions.

After the students complete this pretest, they work on the tutorial. The tutorial begins by guiding students to perform a brief comparative analysis of several other measurement devices (the voltmeter, oscilloscope and spectral analyzer) in addition to the lock-in. This serves to motivate the value of the lock-in for making accurate measurements of the amplitude of a specific frequency within a given signal. This part of the tutorial leads into an in-depth examination of the dual channel lock-in which begins with a short narrated video followed by a series of slides that provide a detailed explanation of a diagram of the dual channel lock-in, including the basic function of the primary components. This brings the students to a mathematical treatment of the effect of the mixer in the lock-in. The student is then guided to find the mathematical expressions for $V_{MX}$ and $V_{MY}$ both in the general case (when $f_S \neq f_R$) and in the most typical case (when $f_S = f_R$). Particular attention is paid to having students focus on which frequencies will be in the unfiltered output signals for each of these two cases to ensure that students internalize both the sum frequency ($f_S + f_R$) and difference frequency ($|f_S - f_R|$). This is followed by having students learn the workings of the low pass filter. The gradual explanation of the low pass filter begins with a diagram of the amplitude loss with respect to the frequency for the low-pass filter for 6 dB/Oct, 12 dB/Oct and 24 dB/Oct rolloff. While learning about these, the students are asked to note two rules of thumb for the 12 dB/Oct rolloff, which is used throughout the remainder of the tutorial. For all frequencies $f \leq 0.1 \times \tau^{-1}$ the students are asked to assume that there is practically no attenuation of the signal as it passes through the filter, while for all frequencies $f \geq 10 \times \tau^{-1}$ the amplitude is strongly attenuated.

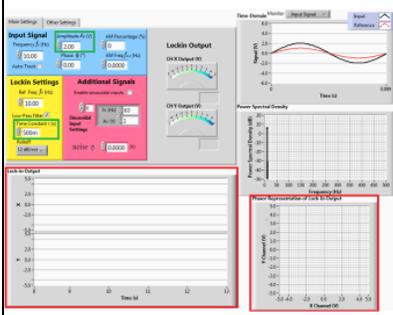

**FIGURE 1.** Example of a typical simulation question (left) and the associated "purpose" slide (right). For the simulation question, red bounding boxes indicate information that the student is expected to sketch or provide, consistent with the lock-in settings and/or output and green bounding boxes highlight settings that were changed from a prior simulation question.

**FIGURE 2.** Example of the explanations provided to students for each simulation (Simulation 2 in this case). The "Explanation" (left) features a conceptual description of the problem's solution while the "Mathematical Explanation" (right) features the mathematics required to solve the problem.

After learning about the basic treatment of the low-pass filter, students move on to make use of the simulation to solidify concepts. The simulation allows the students to manipulate all of the settings commonly found on a lock-in as well as modifying the characteristics of a simulated input signal. The lock-in settings that the student is able to manipulate include the reference frequency, the time constant and the falloff of the low-pass filter. The simulation allows for a considerable degree of control for the input signal. First, the user can specify the frequency and amplitude of the primary input frequency. Additionally, this simulation allows the student to modify the phase of the primary input frequency with respect to the reference frequency. A sinusoidal amplitude modulation can also be introduced to the primary input frequency by specifying the frequency of the modulation and the percentage of the initial amplitude that the primary input frequency amplitude will change by. A secondary frequency can also be added to the input signal to simulate sources of coherent noise and the amplitude and frequency of this second component can be specified. Finally, a white noise can be introduced into the input signal to help demonstrate the effectiveness of the lock-in at measuring the amplitude of a specific frequency despite the presence of white noise. The output of the lock-in simulation is provided to the user in three formats to ensure that the output is available to them in a format that is familiar to them and students are also given opportunity to connect the different forms of output in order to build a good grasp of the workings of this device.

The simulation component of the tutorial begins by familiarizing students with the interface used by the simulation in a similar manner to the pretest supplement. Additionally, students are walked through an example of how to predict the output signal of the lock-in by first applying the equations derived earlier in the mathematical treatment of the mixer and then applying the rules of thumb regarding the low-pass filter. The students then work through a series of problems designed to be used with the simulation similar to the one in Figure 1 (but in different contexts). For each problem, the student is asked to predict the output signal when provided with the input signal and lock-in settings. After they have sketched their predictions, they enter input parameters into the simulation and compare the output signal that the simulation provides with their prediction. After each prediction and simulation, to scaffold student learning, both a mathematical and a more conceptual intuitive explanation of the output signal based on the input parameters are provided as shown in Figure 2, in case the students cannot reconcile their predicted output with the output shown on the screen in the simulation. In Figure 2, the conceptual explanation describes the reasons for the output obtained without explicitly working out the mathematics associated with the mixer or the low-pass filter rules of thumb. The mathematical explanation, on the other hand, provides the mathematics used to determine the pre low-pass filter signal, the low-pass filter rules of thumb and finally the filtered output signal. The overall goal of these scaffolding tools is for students to integrate conceptual and quantitative aspects of the lock-in to develop a functional understanding of the workings of the lock-in.

**FIGURE 3.** An example of one of the tutorial questions which asks students to predict the input signal characteristics (left) and the answer to that question (right).

Following these simulation questions, the student works on a second set of questions that ask them to identify characteristics of either the input signal or the lock-in's settings by examining the provided output signal as shown in Figure 3. The student is asked to either fill in the blank components of the input signal or asked to answer a multiple choice question regarding one of these parameters. Throughout these questions which build on each other, several hints in the form of guiding statements or questions are provided to scaffold student learning (for each of these questions as in Figure 4). One type of hint commonly provided to the student as a first hint takes the form of a hypothetical discussion between two students that are attempting to solve that problem. This hint provides two viewpoints on that problem, one of which is correct and one of which is incorrect, for the student to consider. A more heavily scaffolded second hint commonly takes the form of providing three possible answers with minor differences for the student to consider. This allows the student to consider how each variation in the input signal results in a different output signal. Using a more lightly scaffolded hint first, followed by a more heavily scaffolded hint, allows the student to develop self-reliance as they become less dependent upon the hints (especially the heavily scaffolded hint) as they work through the tutorial, enabling the student's reliance on the scaffolding to fade [22]. After the students have answered each question to the best of their ability, the correct answer is provided along with an explanation for why that answer must be correct (see Figure 3). The guiding problems that build on each other (and become more complex later) allow the student an opportunity to predict the outcome of the experiment in a given situation and determine input parameters and lock-in settings based on the output signal or output parameters for input settings. Once the students have worked on the tutorial, they work on the posttest.

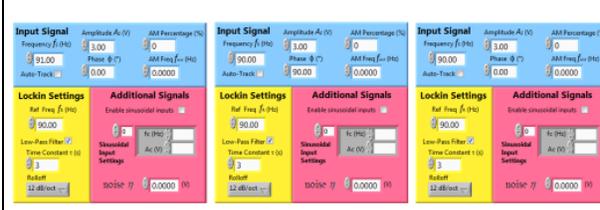

**FIGURE 4.** Examples of two types of hint commonly provided to scaffold student learning.

The posttest is structured in a similar manner to the pretest, consisting of a short quiz comprised of questions that demonstrate the same cases as in the pretest. Students are again allowed to have access to the same supplementary material for the posttest as in the pretest. To ensure that any change in score between the pretest and posttest was solely the result of the tutorial and not a difference in the specific problems used in each quiz, the questions used on the pretest and posttest are switched for roughly half of the students tested.

## STUDENT DIFFICULTIES

Throughout student interviews, many common difficulties that graduate students have in using this device were identified when they were asked to make predictions about either the output signal or input signal of the lock-in in diverse situations. We observed that, before working on the tutorial, most of the interviewed graduate students had an incomplete understanding of the intricacies of the lock-in's operation and many of the interviewed students, even those with extensive experience using the lock-in to make measurements in a lab setting, were only capable of predicting the input or output for the simplest of cases. The difficulties found in our investigation have roots in a lack of coherent understanding of the fundamentals of a lock-in. For example, students often had a fuzzy understanding of what the mixers in a lock-in do. Even in the cases that are most typically encountered in the lab setting, interviewed students often demonstrated only a superficial understanding of the lock-in. The range and prevalence of these difficulties during the interviews demonstrates that students are often using the lock-in as a black box to make measurements at a targeted frequency without understanding the intricacies of its operation. They are unlikely to troubleshoot difficulties as many of the interviewed faculty members had already pointed out during their interviews. The following are difficulties commonly observed among interviewed students before exposure to the tutorial or in the early portions of the tutorial as these difficulties were encountered for the first time.

**Difficulty in calculating and using the corner frequency:** Interviews suggest that one major aspect of the lock-in that is often overlooked by graduate students is the effect of the low pass filter on the output signal. For example, interviewed students had great difficulty with the fact that the frequencies that will make it into the output signal can be estimated by making use of the time constant, τ. Many students ignored the time constant throughout the pretest with some commenting that they didn't know how to account for the time constant. For example, one student stated during a think aloud interview while working through the tutorial that "The most helpful part [of the tutorial] is [when] it explains the [low-pass] filter. That is a part I seldom use in the lock-in". The student went on to explain that he didn't understand how to set the time constant and had little understanding of its purpose in the lock-in despite having worked with the device for roughly a month for authentic research. This confusion regarding the effect of the time constant results in students being unable to properly predict what frequencies will make it through the low-pass filter with little attenuation and thus the frequencies that will appear in the output signal.

**Difficulty with the case in which $f_S = f_R$ and the $2f_R$ signal is strongly attenuated:** Since the lock in amplifier is most commonly used to measure the amplitude of a targeted frequency present in an input signal, the case in which the signal and reference frequency are equal and the time constant is high enough to attenuate the $2f_R$ signal is the most familiar to most of those who have used the lock-in in a lab setting. Despite this being the most common case, some students still had difficulty with predicting the output signal from a lock-in under these conditions. These students most commonly incorrectly thought that the lock-in's output should have a frequency equal to that of the signal frequency. They thought that the lock-in is providing an output similar to that which would be shown by an oscilloscope measuring the input signal. This difficulty demonstrates a misunderstanding of the operation of the lock-in for these conditions. Since the purpose of the lock-in is to measure the amplitude of the input signal at the targeted frequency (determined by the reference frequency) by multiplying the input signal to a DC output, a frequency cannot be present in the output when $f_S = f_R$ and the $2f_R$ signal is strongly attenuated.

**Difficulty with the case in which $f_S \neq f_R$, the $f_R - f_S$ signal is not strongly attenuated and the $f_R + f_S$ signal is strongly attenuated:** This difficulty with the operation of the lock in amplifier affects the student's ability to make predictions regarding other lock-in setups. In the case in which $f_S \neq f_R$ students demonstrated several difficulties rooted in a lack of conceptual understanding of the operation of the lock-in. One common difficulty among students when confronted with this situation involved students stating that the lock-in would not show an output from any frequency in the input signal that does not equal the reference frequency. These students commonly determined that both the x and y output of the lock-in would be zero. It is worth noting that this result was often due to students commonly disregarding the time constant provided in the question statement. For example, one student demonstrated this difficulty when he said "So there's a frequency difference between them (the reference and signal frequency)... So I expect zero for this [output] although the frequencies are close". It appears that the student is considering the effect of the reference and signal frequency being close to one another. Despite this consideration, the student selected zero as the output for both channels. Additionally, this student went on to say "…the entire point of the lock-in is that it should pick out only the frequency… only the component of the same frequency as the reference." a few minutes later in the interview demonstrating that he also thought that the lock-in is only capable of measuring input signals with a frequency equal to the reference frequency.

Another common difficulty among students is that many thought that the lock-in will yield a DC output even under the condition, $f_S \neq f_R$. These students came up with an output that would have been correct if the signal and reference frequency had been equal to each other rather than an output with a frequency of $|f_S - f_R|$ present in it with effectively no attenuation from the low-pass filter. This difficulty is partly due to their lab experience being primarily focused on the most common case. Since some students only make use of the lock-in as a measurement device under the condition that $f_S = f_R$ they have difficulty with any situation that causes the lock-in's output to be anything but a DC signal. Thus, many of these students thought that the output signals from a lock-in must always be a DC signal. Interviews suggest that this type of incorrect rote application is often due to the fact that students are using the lock-in as a black box rather than understanding the operation of this device before making use of it. This type of difficulty can prevent students from troubleshooting if they encounter unexpected output in their lab experiments.

**Difficulty with the case in which $f_S$ and $f_R$ are out of phase ($\varphi \neq 0$):** The most common difficulty that students had with non-zero phase involves cases in which $f_S \neq f_R$, the $f_R - f_S$ signal is not strongly attenuated and the $f_R + f_S$ signal is strongly attenuated. Students would commonly combine factors of the correct solution to this case with factors that are used to solve the case where $f_S = f_R$ and $\varphi \neq 0$. Most commonly, this resulted in a time varying output with different amplitudes for the x-channel and y-channel outputs. Students would often determine the amplitude of the x-channel to be $A_S \cos(\varphi)$ and the amplitude of the y-channel to be $A_S \sin(\varphi)$ rather than $\varphi$ only determining the starting point of the x-channel and y-channel outputs. Additionally, students showed weaker performance on problems in which $f_S = f_R$ and $\varphi \neq 0$.

**Difficulty with the case in which $f_S = f_R$ and the $2f_R$ signal is unattenuated:** Another situation that caused a considerable amount of difficulty among students is when $f_S = f_R$ and the $2f_R$ signal appears in the output. Most commonly students answering this question provide a DC output without a $2f_R$ signal. The difficulty in answering this type of question stems from two sources that have already been mentioned. First, students making use of the lock-in as a black box for the most common case in which they observe a DC output if they chose a sufficiently high time constant when using the device. Additionally, the students' lack of experience with the time constant and how it affects the attenuation of signals in the output signal often leads to them thinking that the low pass filter will remove all frequencies in the output signal.

In addition to these two common difficulties, we find that many students had an even more fundamental issue with the operation of the lock-in that led to difficulty with this case. In particular, many students lacked a coherent understanding of the internal operation of the lock-in and how the multiplication of the input signal and reference signal results in a sum and difference frequency in the output signal. Most of the students interviewed showed an understanding of the difference frequency which resulted in a DC output in this case but failed to even mention the existence of the sum frequency which would result in the $2f_R$ signal in this case. Additionally, none of the students interviewed made use of the relevant equations for the multiplication of the input and reference signals in the mixer to gain insight into what output they should expect when the questions were posed in the pretest.

**Difficulty with amplitude modulation of the input signal:** Another case in which students had difficulty during the think aloud interview is the introduction of an amplitude modulation to the primary input frequency. The most common difficulty observed during interviews is that students often thought that the amplitude modulation would affect both the x-channel and y-channel output signal even when the amplitude only affects the output in one of the two channels. Interviews suggest that this difficulty is due to a combination of a lack of experience with amplitude modulation paired with the students' experience with other cases that introduce frequencies into the output signal. For all other cases that result in frequencies in the output signal of the lock-in, the frequency appears in both channels of the output with equal amplitude. This prior experience colors the students expectations about the possible output signal of the lock-in. Many students also had difficulty determining how the low-pass filter affects the amplitude modulation. Despite amplitude modulation generating a frequency that is passed through the low-pass filter in the same manner as any other frequency, some students either thought that amplitude modulation should not be affected by the low-pass filter or completely ignored consideration of the effects of the low-pass filter on this signal.

**Difficulty with multiple frequencies present in the input signal:** One final aspect of the lock-in that practically all interviewed students had difficulty with was the case in which multiple frequencies are present in the input signal. Interviews suggest that many students had not considered such a possibility in the past. Although the difficulty with this case was common to all students interviewed, their reactions to this problem during interviews were often varied with only one common type of student response. Most commonly students ignored one of the two frequencies and continued to treat the other frequency as though it was the only one in the input signal. One student demonstrated this confusion with the statement "So, since I don't really know how to incorporate the changing (indicates the additional signals section) into it, I'm just going to stick with the amplitude and phase (indicates the primary input signal section)". Thus, when this student faced a situation that he did not have experience with, he opted to ignore the source of confusion rather than find a way to incorporate it into the output signal. This tendency can have a detrimental effect on student ability to troubleshoot as was already pointed out by the interviewed faculty members in actual situations in research labs.

## RESULTS FROM THE PRE-/POSTTEST

**TABLE 2.** Summary of the average grade, standard deviation and number of instances of each of these cases present in both the pretest and posttest.

|  | Average Pretest Score | Pretest Standard Deviation | Number of Instances in Pretest | Average Posttest Score | Posttest Standard Deviation | Number of Instances in Posttest | p-value |
|---|---|---|---|---|---|---|---|
| Case 1 | 76.4% | 40.6% | 34 | 100% | 0.0% | 33 | 0.002 |
| Case 2 | 22.0% | 34.7% | 33 | 90.9% | 21.2% | 33 | <0.001 |
| Case 3 | 55.9% | 41.3% | 36 | 89.2% | 24.7% | 30 | <0.001 |
| Case 4 | 33.0% | 32.3% | 22 | 68.2% | 30.3% | 22 | <0.001 |
| Case 5 | 26.2% | 32.3% | 21 | 83.0% | 34.9% | 22 | <0.001 |
| Case 6 | 25.0% | 38.4% | 33 | 80.3% | 33.6% | 33 | <0.001 |

As noted earlier, to evaluate the effectiveness of the tutorial at addressing these difficulties, a pretest and posttest were given to all students who worked on the tutorial. The difficulties discussed in the previous section are primarily determined by examining student performance on the pretest as well as the tutorial and asking students for clarification of the points they had not made clear by themselves during the individual interviews in which they first worked on the pretest, then they worked on the tutorial and then took the posttest.

In order to examine students understanding, six potential configurations of lock-in settings were included in the pretest and posttest questions. In the following section, each of these cases will be discussed in detail along with an examination of the tutorial's effectiveness at addressing student difficulties related to each case. During the development and evaluation of the tutorial, a few additional pretest and posttest questions were included in the later version. This accounts for the varying number of instances of a particular case present in the pretest and posttest shown in Table 2.

**(Case 1) $f_R = f_S$ with the $2f_R$ signal strongly attenuated and $\varphi = 0$:** This case is the most commonly encountered lab situation involving the lock-in in which students are attempting to measure the amplitude of the frequency $f_S$ in the input signal and have correctly set the reference frequency equal to $f_S$ and the time constant high enough to filter out the $2f_R$ signal. Despite this case being one that should be typical to students who commonly make use of this device, some students had difficulty with these problems. Examining Table 2, we can see that the average score across all pretest problems that demonstrate this case is roughly 75%. This not-perfect performance results from some students having only limited experience with the lock-in as well as some experienced students who thought that the lock-in would provide an output with frequency $f_R$ in this case as already discussed earlier. While these pretest results are not particularly disheartening, as shown in Table 2, the tutorial does show a marked improvement in student performance with every student that made use of the tutorial answering every question on the posttest that demonstrated this case perfectly (100%). This demonstrates the effectiveness of this tutorial at ensuring that all students understand how the lock-in operates in this simple case.

**(Case 2) $f_R \neq f_S$, the $f_R - f_S$ signal is not strongly attenuated and the $f_R + f_S$ signal is strongly attenuated**: This case is less commonly encountered by students in the lab setting. This situation could arise if the student meant to measure the amplitude of frequency $f_S$ in the input signal by setting $f_R = f_S$ and incorrectly sets $f_R$ to a value near $f_S$. This situation also serves to illustrate the importance of integrating conceptual understanding with mathematical preliminaries that are central to the lock-in's operation. By correctly answering questions that demonstrate this case, students have to apply both the correct general equations for the unfiltered output signal as well as the effect of the low-pass filter for the time constant and rolloff designated for the problem. Examining Table 2, we can see a distinct lack of student understanding on the pretest with the average student score being only slightly over 20% across all pretest problems that demonstrated this case. This lack of understanding is based partly in the interviewed students only rarely encountering this case. Throughout the pretest, most students were only able to correctly answer problems that were similar to the most common case. Student responses in interviews illustrate that students use the lock-in as a black box rather than by identifying the output signal based on a deep understanding of the internal operation of the lock-in. After working through the tutorial, student performance improved to an average of roughly 90%. This improvement was primarily the result of improved student understanding of the mixer as well as the use of the rules of thumb to deal with the low-pass filter.

**(Case 3) $f_R = f_S$ with the $2f_R$ signal strongly attenuated and $\varphi \neq 0$:** This case is very similar to case 1 (the most commonly faced lab situation) with the only difference being that the phase angle is no longer set to zero. This case could commonly be faced by students in the lab setting when the frequency of interest, $f_S$, is out of phase with the reference frequency. This case also served to illustrate the importance of having both an x-channel and y-channel output, which is an issue that was confusing to several students during interviews before they worked on the tutorial. This case shows that the amplitude of the frequency, $f_S$, cannot, in general, be measured completely in either the x-channel or y-channel without the reference frequency being in phase with $f_S$. Students showed decreased performance on these problems when compared to case 1 scoring slightly over 55% on average on the pretest. This is partly the result of students having somewhat less experience with this case than with the case in which $f_S$ is in phase with $f_R$, due to students commonly having control over the phase of the reference frequency in most lab experiments involving the lock-in. This average was improved to nearly 90% after working on the tutorial

**(Case 4) $f_R = f_S$ with the $2f_R$ signal not strongly attenuated and $\varphi = 0$**: This case is primarily experienced by students when they are measuring a frequency, $f_S$, that is very small and the time constant is incorrectly set resulting in the $2f_R$ signal being passed by the low-pass filter. This case is used to demonstrate the importance of correctly accounting for the effect of the low-pass filter and selecting a time constant that will properly limit the frequencies that can pass. Students had little experience with this case before the tutorial leading to a very low average pretest score of under 35.0%. The actual student understanding of the $2f_R$ signal passing through the filter is even lower than the average pretest score which suggests with most of the credit earned on problems that demonstrate this case come from correctly determining the DC offset of the output in accordance with Rubric 3 in Table 1. The tutorial was able to improve student understanding such that the average posttest score was roughly 70%. The relatively lower improvement demonstrated in this case when compared to the others is partly the result of the fact that in all other cases the larger of the two frequencies generated by the multiplier is strongly attenuated while passing through the low pass filter. This leads to students commonly disregarding the effects of the $f_R + f_S$ term on the output signal.

**(Case 5) Amplitude modulation present in the input signal:** Amplitude modulation in the form of a sinusoidal time dependent variance in the amplitude of the input signal is used in this tutorial to examine the effects of a time dependent amplitude on the output of the lock-in. Additionally, this case serves to demonstrate how the value of the time constant can affect the time required to take an accurate measurement. Since the output signal changes more slowly as the time constant increases the time required for a stable output to be achieved in the most common case (case 1) will also increase as the time constant increases. This is important for students making use of the lock-in to make a large number of measurements to ensure that they select a time constant that is high enough to filter out any time varying components yet low enough to allow quick measurements to be taken. The average score on all pretest questions that demonstrate this case is slightly over 25%. This weak performance demonstrates the student's limited understanding of the low-pass filter as well as how a change in the input signal's amplitude would affect the output signal. Students showed a marked improvement in performance after working on the tutorial with the average score being over 80% on posttest questions demonstrating this case.

**(Case 6) Two frequencies present in the input signal:** The final case involves the presence of two frequencies in the input signal with the $f_R - f_S$ component of each signal passing through the low-pass filter not strongly attenuated while both $f_R + f_S$ signals are strongly attenuated. This case serves to demonstrate to students how the lock-in processes multiple frequencies that are present in the input signal to generate a single output signal. It also acts as an example of the potential lab scenario in which the experimenter is attempting to measure the amplitude of one frequency (as in the most common case) but incorrectly sets the time constant too low, allowing part of the mixed signal from a coherent noise with a frequency close to the value of the reference frequency to pass through the low-pass filter. The average student score on the pretest for all problems that demonstrate this case is 25%. This low average score partly illustrates the students' inexperience with both cases involving two frequencies in the input signal and cases with frequencies in the input signal that do not equal the reference signal (similar to case 2). After students work on the tutorial, they obtain an average score of roughly 80% across all posttest questions that demonstrate this case.

Examining the pretest and posttest results of the six cases covered suggests that the tutorial is effective at improving student understanding in a variety of situations ranging from the most commonly experienced lab setting and other less commonly faced cases that illustrate situations that the student may potentially face while making use of this device. Despite the wide variety of experience levels among students who took the pretest and posttest there was little correlation between experience level and pretest scores with highly experienced student only performing slightly better on case 1 problems and all students performing comparably on all other cases. In addition to being based on a variety of potential experimental setups, the pretest and posttest questions can also be divided into two categories based upon what form the answer takes. Questions either asked student to sketch the output signal that should result from a given set of input parameters or ask the student to fill in the blanks about characteristics of the input signal based on the lock-in's settings and a provided output signal diagram [21].

**TABLE 3.** Summary of the average scores, standard deviation and number of instances of predicting input and predicting output questions as well as the total average score on both the pretest and posttest.

|  | Average Pretest Score | Pretest Standard Deviation | Number of Instances in Pretest | Average Posttest Score | Posttest Standard Deviation | Number of Instances in Posttest | p-value |
|---|---|---|---|---|---|---|---|
| Predicting Output | 38.1% | 42.6% | 135 | 87.7% | 27.7% | 151 | <0.001 |
| Predicting Input | 42.5% | 42.1% | 55 | 80.9% | 30.5% | 55 | <0.001 |
| Total Grade | 39.4% | 42.5% | 190 | 85.9% | 28.7% | 206 | <0.001 |

Questions in which the student was asked to predict and sketch the output signal were included to examine the student's ability to check the output signal to ensure that the input parameters are as they predicted. If the students are able to correctly answer problems of this type, they should be able to confirm that the output signal is as they predicted when performing a measurement in the lab and if this prediction does not match their measurement they can determine that the input signal has characteristics that they did not take into account. This ability to predict the form of the output signal before making a measurement should ensure that students are less likely to record anomalous outputs based on an incorrect understanding of the input signal that they are analyzing. Students initially showed a limited ability to predict the output signal for the array of cases that were treated in the pretest. Table 3 shows that this resulted in an average score of under 40% across all pretest questions that had the student sketch the output signal. This average improves to over 85% after students made use of the tutorial.

The remaining questions, which ask students to predict the characteristics of the input signal based on the output signal and the lock-in's settings, evaluate the students' ability to examine an output signal and determine what the input signal must be. This ability is valuable to students in the lab setting when they are faced with an output signal that differs from what they expected. Table 3 shows that student performance on these questions before exposure to the tutorial is roughly 40% across all pretest questions that asked the student to predict the input settings. This average improved to over 80% on all posttest material as shown in Table 3. Improved student performance on either of these two types of problems should correlate with increased student ability to troubleshoot difficulties that may arise when making use of the lock-in in their authentic research scenarios.

## SUMMARY

We have developed and evaluated a research-based tutorial that helps students learn the basics of how this instrument operates and also helps students make connections between the relevant concepts and the pertinent mathematics that describes the operations of its major components. This tutorial helps students predict output signals and input signals in a variety of cases all of which mimic situations that may be faced by the student in the lab setting.

We find that physics graduate students who use lock-ins for their experimental research have many common difficulties with the basic functions of this instrument. These difficulties make it difficult for students to correctly make predictions about the output signal given the input parameters or input parameters given the output signal in any but the most common case with any degree of accuracy. Interviews suggest that the understanding of majority of students is usually limited to the use of the lock-in as a black box to perform measurements which makes it extremely difficult for them to be able to troubleshoot problems (sometimes due to spurious signals that they may not be aware of or choosing incorrect time constant).

Examination of the average student scores on the pretests and posttests shows considerable improvement for all cases discussed. This improved ability to identify both typical (case 1) and less common (case 2-6) characteristics should improve student ability to both ensure that the output signals obtained are as expected and troubleshoot anomalous output signals or unexpected conditions present in the input signal when they arise. Additionally, this increase in score on the posttest across a wide variety of students with varied levels of experience with the lock-in demonstrates the tutorial's utility in both providing an OnRamp for student' who are just being introduced into the lab setting as well as providing an opportunity for those who have been using the lock-in to improve their understanding of this powerful device

## ACKNOWLEDGEMENTS

We thank the National Science Foundation for award NSF-1124131.